\documentclass[%
twocolumn,
preprintnumbers,
 amsmath,amssymb,
 aps,
 american
 prd,
 superscriptaddress,
]{revtex4-1}

\usepackage{amsmath}
\usepackage[english]{babel}
\usepackage{graphicx}% Include figure files
\usepackage{dcolumn}% Align table columns on decimal point
\usepackage{bm}% bold math
\usepackage{dsfont}
\usepackage{slashed}
\usepackage{hyperref}% add hypertext capabilities
\hypersetup{pdfauthor={Name}}
\usepackage{color}

\usepackage[paperwidth=210mm,paperheight=297mm,centering,hmargin=1.25cm,vmargin=2.0cm]{geometry}
\begin{document}

\title{$T-\mu$ quark matter phase transitions and critical end point in nonlocal PNJL models under a strong magnetic field}
\author{J.P. Carlomagno}
\affiliation{IFLP, CONICET departamento de Física, Facultad de Ciencias Exactas, Universidad Nacional de La PLata, C.C. 67, (1900) La Plata, Argentina}
\affiliation{CONICET, Rivadavia 1917, (1033) Buenos Aires, Argentina }
\author{S.A. Ferraris}
\affiliation{Physics Department, Comisión Nacional de Energía Atómica (C.N.E.A), Avenida del Libertador 8250, (1429) Buenos Aires, Argentina}
\author{D. Gómez Dumm}
\affiliation{IFLP, CONICET departamento de Física, Facultad de Ciencias Exactas, Universidad Nacional de La PLata, C.C. 67, (1900) La Plata, Argentina}
\affiliation{CONICET, Rivadavia 1917, (1033) Buenos Aires, Argentina }
\author{A.G. Grunfeld}
\affiliation{CONICET, Rivadavia 1917, (1033) Buenos Aires, Argentina }
\affiliation{Physics Department, Comisión Nacional de Energía Atómica (C.N.E.A), Avenida del Libertador 8250, (1429) Buenos Aires, Argentina}
\date{\today}

\begin{abstract}
We study the $T-\mu$ phase diagram of quark matter under the influence of a strong uniform magnetic field in the framework of a nonlocal extension of the Polyakov–Nambu–Jona-Lasinio model (PNJL). The existence of a critical end point (CEP) is found for the whole considered range of the magnetic field (up to 1~GeV$^2$). We analyze the location of this CEP as a function of the external field and discuss the presence of inverse magnetic catalysis for nonzero chemical potentials. Our results show that the temperature of the CEP decreases with the magnetic field, in contrast to the behavior observed in local NJL/PNJL models. 
\end{abstract}

%\keywords{Suggested keywords}%Use showkeys class option if keyword display desired

\maketitle
\section{Introduction}
The study of the QCD phase diagram under the influence of a strong external
magnetic field has captivated significant attention in recent
years~\citep{Kharzeev:2012ph,Andersen:2014xxa,Miransky:2015ava}. This is due
in part to the fact that the subject has applications in various fields such
as cosmology, heavy ion collisions and compact stars physics. In fact, it
has been argued~\citep{Vachaspati:1991nm} that extremely high magnetic
fields (reaching orders of magnitude of about $10^{20}$~G) could have
existed during the cosmological electroweak phase transition; in addition,
in ultra-peripheral heavy-ion collisions the generated magnetic fields are
shown to be proportional to the collision energy, reaching up to $\sim 5
\times 10^{19}$~G~\citep{Deng:2012pc}, while in an astrophysical scenario,
the magnetic field on the surface of magnetars is estimated to be of
the order of $10^{14}$~G~\citep{Duncan:1992hi}, and significantly higher
values are expected to be found in the deep interior of these objects. Since
the order of magnitude of such huge magnetic fields is comparable to the QCD
confining scale squared ($|eB|\agt \Lambda_{\rm QCD}^2$ for $|B|\agt
10^{19}$~G), the above scenarios offer interesting opportunities to probe
the QCD phase diagram in the region of deconfinement and chiral symmetry
restoration transitions.

Due to the nonperturbative character of strong interactions in this regime
of interest, one is lead either to use lattice QCD (LQCD) techniques or to
rely on effective theories that could capture the main features of QCD
phenomenology showing at the same time consistency with LQCD results. Among
these effective theories, the Nambu-Jona-Lasinio (NJL)
model~\citep{Nambu:1961tp,Nambu:1961fr} is one of the most widely used
approaches. In this effective chiral quark model, quarks interact through
local four-point vertices, leading to the spontaneous breakdown of chiral
symmetry for a sufficiently large coupling
strength~\citep{Vogl:1991qt,Klevansky:1992qe,Hatsuda:1994pi}. In fact, it
has been pointed out that the model can be improved by considering nonlocal
separable interactions, which arise naturally in several effective
approaches to QCD and lead to a better agreement with LQCD
calculations~\citep{Schmidt:1994di,Burden:1996nh,Bowler:1994ir,Ripka:1997zb}.
A recent review that discuss the features of nonlocal NJL models, including
the analysis of strong interaction matter under extreme conditions, can be
found in Ref.~\citep{Dumm:2021vop}.

In this work we study the hadron-quark QCD phase diagram in the context of a
nonlocal two-flavor NJL
model~\citep{General:2000zx,GomezDumm:2001fz,dumm2006covariant} that also
includes the interaction between the quarks and the Polyakov
loop~\citep{tHooft:1977nqb,Polyakov:1978vu,Meisinger:1995ih,Fukushima:2003fw,
Megias:2004hj,Ratti:2005jh,Roessner:2006xn,Mukherjee:2006hq,Sasaki:2006ww}.
This so-called nonlocal Polyakov-Nambu-Jona-Lasinio model
(nlPNJL)~\citep{Contrera:2007wu,Hell:2008cc,Carlomagno:2013ona} provides an
order parameter for the confinement/deconfinement transition, and leads to a
critical temperature for chiral symmetry restoration that is found to be in
good agreement with LQCD results~\citep{Karsch:2003jg}. Noticeably, nonlocal
PNJL models show the interesting feature of being able to reproduce in a
naturally way the effect known as inverse magnetic catalysis
(IMC)~\citep{Pagura:2016pwr,dumm2017strong}: while at zero temperature the
presence of a strong magnetic field leads to an enhancement of the
quark-antiquark chiral condensate (favouring, or ``catalizing'', the
breakdown of chiral symmetry), this behaviour becomes the opposite if the
temperature gets significantly increased. The observation of IMC from LQCD
calculations~\citep{Bali:2011qj,Bali:2012zg} represents a challenge for most
naive effective approaches for low energy QCD (including the usual local NJL
model), in which this effect is not
found~\citep{Andersen:2014xxa,Kharzeev:2012ph,Miransky:2015ava}.

In the framework of nlPNJL models, we concentrate here on the study of the
deconfinement and chiral restoration transitions in the $T-\mu$ phase
diagram (where $T$ is the temperature and $\mu$ the quark chemical
potential), considering the presence of an external uniform and static
magnetic field $\vec B$. This has been previously addressed in the context
of the Ginzburg-Landau formalism, the linear sigma model, and various versions of local NJL-like models, see
Refs.~\citep{Ruggieri:2014bqa,Inagaki:2003yi,Avancini:2012ee,Ferrari:2012yw,Costa:2013zca,Costa:2015bza,
Rechenberger:2016gim,Ferreira:2017wtx,Tawfik:2017cdx,Ferreira:2018pux,Ayala:2021nhx}.
Our analysis within the nonlocal PNJL approach can be viewed as an extension
of two previous works, Refs.~\citep{dumm2017strong}
and~\citep{Ferraris_2021}, in which systems at nonzero temperature and
nonzero density were separately considered. A previous analysis has been also done within the nonlocal NJL, restricted to the weak magnetic field limit~\citep{Marquez:2017uys}. Besides the behaviour of
quark-antiquark condensates in the $B-T-\mu$ parameter space, in the present work we concentrate
on the features of deconfinement and chiral restoration transition critical
lines, focusing in the position of the critical endpoint (CEP) that separates the first-order transition line from the crossover region. We believe that
the location of this point, and, in particular, its dependence with the
magnetic field, is an interesting matter of discussion. Given the
difficulties in getting accurate estimations from LQCD for large values of
$\mu$ (owing to the well-known sign problem), it is worth studying the
comparison between the predictions obtained from different effective models
for strong interactions. In this sense, we notice that our results differ
qualitatively from those obtained within the local NJL/PNJL
approaches~\citep{Avancini:2012ee,Ferrari:2012yw,Costa:2013zca,Costa:2015bza}. In our work
we also analyze the dependence of the results on the model parametrization,
discussing the compatibility with the existence of IMC.

The article is organized as follows. In Sec.~II we describe the theoretical
formalism for magnetized quark matter within the nlPNJL model at nonzero
temperature and chemical potential. Next, in Sec.~III we show our numerical
results for the phase transition order parameters and critical lines,
considering different strengths of the magnetic field. Finally, in Sec.~IV
we provide a summary and conclusions of our work.

\section{formalism}
We start by defining the effective Euclidean action in our nonlocal NJL
model for two quark flavors $u$ and $d$,
\begin{equation}
S_{E}=\int d^{4}x \left[\bar{\psi}(x)\left(-i\slashed{\partial} + m_{c}\right)\psi(x)-\frac{G}{2}j_{a}(x)j_{a}(x)\right]\ ,
\label{accionE}
\end{equation}
where $\psi$ stands for the quark field doublet and $G$ is a coupling
constant. We assume that the current quark mass $m_{c}$ is the same for both
flavors, $m_{c}=m_{u}=m_{d}$. The nonlocal currents $j_{a}(x)$ are defined
as
\begin{equation}
j_{a}(x)=\int d^{4}z\ \mathcal{G}(z)\,\bar{\psi}\left(x + \frac{z}{2}\right)\Gamma_{a}\psi\left(x - \frac{z}{2}\right)\ ,
\label{corrientes}
\end{equation}
where $\Gamma_{a}=(\mathds{1},i\gamma_{5}\vec{\tau})$ and $\mathcal{G}(z)$
is a form factor. In order to introduce the interaction with an external
magnetic field $\vec{B}$ one has to replace the partial derivative
$\partial_{\mu}$ in the kinetic term of the effective action in
Eq.~(\ref{accionE}) by a covariant derivative, i.e.
\begin{equation}
\partial_{\mu} \ \rightarrow \ D_{\mu}\equiv \partial_{\mu}-i\hat{Q}\mathcal{A}_{\mu}\ ,
\end{equation}
where $\mathcal{A}_{\mu}$ stands for the external electromagnetic field, and
$\hat{Q}={\rm diag}(q_{u}, q_{d})$, with $q_{u}=2e/3$, $q_{d}=-e/3$, is the
electromagnetic quark charge operator. In order to preserve gauge
invariance, this replacement also implies a change in the nonlocal currents
in Eq.~(\ref{corrientes}) given by
\citep{dumm2006covariant,noguera2008nonlocal}
\begin{gather}
\psi(x-z/2) \ \rightarrow \ \mathcal{W}(x,x-z/2)\,\psi(x-z/2)\ ,\nonumber\\
\psi(x+z/2)^{\dagger} \ \rightarrow \ \psi(x+z/2)^{\dagger}\,\mathcal{W}(x+z/2,x)\ ,
\end{gather}
where the function $\mathcal{W}(s,t)$ is defined as
\begin{equation}
\mathcal{W}(s,t)=P\,\exp\left[-i\int_{s}^{t} dr_{\mu}\hat{Q}\mathcal{A}_{\mu}(r) \right]\ .
\label{Funcw}
\end{equation}
As it is usually done~\citep{bloch1952field}, in the above integral we take
a straight line path connecting $s$ with $t$.

To proceed, we consider the case of a static and uniform external magnetic
field oriented along the 3-axis. We choose to work in the Landau gauge,
taking $\mathcal{A}_{\mu}=Bx_{1}\delta_{\mu 2}$. Then the function
$\mathcal{W}(s,t)$ defined in Eq.~(\ref{Funcw}) is given by
\begin{equation}
\mathcal{W}(s,t)=\exp\left[-\frac{i}{2}\hat{Q}B\left(s_{1} + t_{1}\right)\left(t_{2} - s_{2}\right) \right]\ .
\label{Funcw2}
\end{equation}
Since the degrees of freedom of quark fields are not observed at low
energies, the fermions can be integrated out, and the action can be written
in terms of scalar and pseudo-scalar fields $\sigma(x)$ and $\vec{\pi}(x)$,
respectively. This is a standard procedure that leads to the bosonized
action~\citep{noguera2008nonlocal,gomez2011pion,GomezDumm:2017jij}
\begin{align}
S_{\rm bos} = &  -\ln\,\det\mathcal{D}_{x,x'} \nonumber \\
& + \frac{1}{2G}\int d^{4}x \left[\sigma(x)\sigma(x)+\vec{\pi}(x)\cdot\vec{\pi}(x) \right],
\label{Sbos}
\end{align}
where
\begin{align}
\mathcal{D}_{x,x'} = &~ \delta^{(4)}(x-x')\left(-i\slashed{D}+m_{c}\right) + \mathcal{G}(x-x')\gamma_{0}\nonumber\\
&\times\mathcal{W}(x,\bar{x})\gamma_{0}\left[\sigma(\bar{x}) + i\gamma_{5}\vec{\tau}\cdot\vec{\pi}(\bar{x})\right]\mathcal{W}(\bar{x},x'),
\end{align}
with $\bar{x}=(x+x')/2$.

We consider now the mean field approximation (MFA). Thus, we expand the
mesonic fields in terms of the corresponding vacuum expectation values and
their fluctuations, $\sigma(x)=\bar{\sigma} + \delta \sigma(x)$ and
$\vec{\pi}(x)=\delta \vec{\pi}(x)$. Spontaneous breakdown of chiral symmetry
leads to a nonzero translational invariant vacuum expectation value
$\bar{\sigma}$ for the $\sigma(x)$ field, while for symmetry reasons the
mean field values for the pseudoscalars are $\pi_a(x)=0$. In this way the
MFA bosonized action per unit volume can be written as
\begin{equation}
\frac{S_{\rm bos}^{\rm MFA}}{V^{(4)}}=\frac{\bar{\sigma}^{2}}{2G} - \frac{N_{c}}{V^{(4)}}\sum_{f=u,d} {\rm tr}\,\ln\left[\mathcal{D}_{x,x'}^{{\rm MFA},f}\right],
\label{S_bos_MFA_Vol}
\end{equation}
where $N_{c}$ is the number of colors.
%To calculate the traces over Dirac and coordinate spaces, it is convenient to perform a Ritus transform of $\mathcal{D}_{x,x'}^{{\rm MFA},f}$ \citep{ritus1978method}.

Next, using the standard Matsubara formalism we extend our analysis to a
system at both nonzero temperature $T$ and nonzero chemical potential $\mu$.
As stated, the cases of finite $T$ and $\mu$ have been separately considered
in previous works, see Refs.~\citep{dumm2017strong} and
\citep{Ferraris_2021}. The purpose of this article is to study the combined
effect of both thermodynamic variables in the system. To account for the
confinement/deconfinement effects we also include the coupling of fermions
to the Polyakov loop (PL). We assume that quarks move in a constant color
background field $\phi=ig\delta_{\mu0}G_{a}^{\mu}\lambda^{a}/2$, where
$G_{a}^{\mu}$ are color gauge fields. It is convenient to work in the
so-called Polyakov gauge, in which the matrix $\phi$ is given in a diagonal
representation $\phi = \phi_{3}\lambda_{3} +\phi_{8}\lambda_{8}$ with only
two independent variables, $\phi_{3}$ and $\phi_{8}$. Then the traced
Polyakov loop $\Phi=\frac{1}{3}{\rm Tr}\,\exp(i\phi/T)$ can be used as an
order parameter for the confinement/deconfinement transitions. Owing to the
charge conjugation properties of the QCD Lagrangian~\citep{dumitru2005dense}
one expects $\Phi$ to be real, which implies
$\phi_{8}=0$~\citep{Roessner:2006xn} and $\Phi=[1+2\cos(\phi_{3}/T)]/3$.
Finally, to describe color gauge field self-interactions we also include in
the Lagrangian an effective potential $\mathcal{U}(\Phi,T)$. Here we take
for this potential a form based on a Ginzburg-Landau
ansatz~\citep{ratti2006phases,scavenius2002k}
\begin{equation}
\frac{\mathcal{U}(\Phi,T) }{T^{4}} = -\frac{b_{2}(T)}{2}\Phi^{2} - \frac{b_{3}}{3}\Phi^{3}+\frac{b_{4}}{4}\Phi^{4}\ ,
\end{equation}
where
\begin{equation}
b_{2}(T) = a_{0}+a_{1}\left(\frac{T_{0}}{T} \right) + a_{2}\left(\frac{T_{0}}{T}\right)^{2} + a_{3}\left(\frac{T_{0}}{T} \right)^{3}\ .
\end{equation}
Following Ref.~\citep{ratti2006phases}, for the above coefficients we take
the numerical values $a_{0}=6.75$, $a_{1}=-1.95$, $a_{2}=2.625$,
$a_{3}=-7.44$, $b_{3}=0.75$ and $b_{4}=7.5$. For the reference temperature
$T_0$ we take the value 210~MeV, which has been found to be adequate for the
case of two light dynamical quarks~\citep{Schaefer:2007pw}.

Under the above assumptions, after some calculation we can obtain the grand
canonical thermodynamic potential for a system at finite temperature $T$ and
chemical potential $\mu$ in the presence of the external magnetic field. We
get
\begin{align}
\Omega_{B,T,\mu}^{\rm MFA} = & \frac{\bar{\sigma}^{2}}{2G} - T\sum_{n=-\infty}^{\infty}
\sum_{c=r,g,b} \sum_{f=u,d} \frac{|q_{f}B|}{2\pi} \nonumber \\
& \times \int\frac{dp_{3}}{2\pi}\,
\bigg\{\ln\left[p_\parallel^{2} + \left( M_{0,p_\parallel}^{\lambda_{f},f} \right)^{2} \right] \nonumber \\
& + \sum_{k=1}^{\infty}\,\ln\Delta_{k,p_\parallel}^{f} \bigg\}  + \mathcal{U}(\Phi,T)\ ,
\label{granpotMFA}
\end{align}
where
\begin{align}
\Delta_{k,p_\parallel}^{f}=& \left( 2k|q_{f}B|+p_\parallel^{2} + M_{k,p_\parallel}^{+,f}M_{k,p_\parallel}^{-,f}\right)^{2}\nonumber\\
& +p_\parallel^{2}\left(M_{k,p_\parallel}^{+,f} -M_{k,p_\parallel}^{-,f}\right)^{2}\ .
\end{align}
Here the functions $M_{k,p_\parallel}^{\pm,f}$ are given by
\begin{equation}
M_{k,p_\parallel}^{\lambda,f} = (1-\delta_{k,-1})\,m_{c} +
\bar{\sigma}\,g_{k,p_\parallel}^{\lambda,f}\ ,
\label{masa_const}
\end{equation}
where
\begin{align}
g_{k,p_\parallel}^{\lambda,f}=&\frac{4\pi}{|q_{f}B|}\left(-1\right)^{k_{\lambda}}
\int \frac{d^{2}p_{\bot}}{\left(2\pi\right)^{2}}\ g(p_{\bot}^{2} + p_\parallel^{2})
\nonumber \\
& \exp\left(-p_{\bot}^{2}/B_f\right)L_{k_{\lambda}}(2p_{\bot}^{2}/B_f)\ ,
\label{funcg}
\end{align}
$L_n(x)$ being the Laguerre polynomials. The function
$g(p_\perp^2+p_\parallel^2)$ is the Fourier transform of the nonlocal form
factor ${\cal G}(z)$ in Eq.~(\ref{corrientes}). In the above expressions we
have used the definition $k_{\pm}=k-1/2\pm s_{f}/2$, where $s_{f}={\rm
sign}(q_{f}B)$ and $k$ stands for the so-called Landau level, while
$\omega_{n}=(2n+1)\pi T$ are the Matsubara frequencies corresponding to
fermionic modes. We distinguish a two-momentum vector $\vec p_\bot$
perpendicular to the magnetic field, and a vector $\vec p_\parallel \equiv
(p_{3}, \omega_{n}-i\mu-\phi_{c})$. The subscripts $c=r,g,b$, and $f=u,d$
stand for color and flavor indices, respectively, and color background
fields are given by $\phi_{r}=-\phi_{g}=\phi_{3}$, $\phi_{b}=0$. Notice that
the expression in Eq.~(\ref{masa_const}) can be interpreted as a constituent
quark mass, which depends on the external magnetic field and is a function
of the momentum due to the nonlocal character of the four-fermion
interactions.

The integral in Eq.~(\ref{granpotMFA}) is divergent and has to be
regularized. We use a prescription often considered in the
literature~\citep{dumm2005characteristics}, in which a ``free'' term is
subtracted and then added in a regularized form, namely
\begin{equation}
\Omega_{B,T,\mu}^{\rm MFA, reg}=\Omega_{B,T,\mu}^{\rm MFA} -
\Omega_{B,T,\mu}^{\rm free} + \Omega_{B,T,\mu}^{\rm free,reg}.
\end{equation}
Here the ``free'' piece $\Omega_{B,T,\mu}^{\rm free}$ is evaluated at
$\bar{\sigma}=0$, but keeping the interaction with the magnetic field and
the Polyakov loop. The regularized term $\Omega_{B,T,\mu}^{\rm free,reg}$
is given by~\citep{Menezes:2008qt}
\begin{align}
\Omega_{B,T,\mu}^{\rm free,reg}= & -\frac{3}{2\pi^{2}}\sum_{f}\left(q_{f}B\right)^{2}\left[\zeta^{'}(-1,x_{f}) + F(x_{f}) \right]\nonumber\\
&-T\sum_{c,f,k} \frac{|q_f B|}{2\pi}~\alpha_k \int \frac{dp}{2\pi}~
G_{k,p}^{f}(\phi_{c},\mu,T)\ ,
\end{align}
where
\begin{align}
&F(x_{f}) =\nonumber \frac{x_{f}^{2}}{4}-\frac{1}{2}\left(x_{f}^{2} - x_{f}\right)\ln(x_{f})\ , \\
&\nonumber G_{k,p}^{f}(\phi_{c},\mu,T)=\sum_{s=\pm} \ln \left\{ 1 + \exp[-( \epsilon_{kp}^f +i \phi_{c} + s \mu) /T ]\right\}\ .
\end{align}
Here we have used the definitions $x_{f}=m_{c}^{2}/(2|q_{f}B|)$,
$\alpha_{k}=2-\delta_{k,0}$, $\epsilon_{kp}^{f}=\left( (2k|q_{f}B|) +p^{2}
+m_{c}^{2}\right)^{1/2}$ and
$\zeta^{'}(-1,x_{f})=d\zeta(z,x_{f})/dz|_{z=-1}$, where $\zeta(z,x_{f})$ is
the Hurwitz zeta function.

Finally, $\bar{\sigma}(B,T,\mu)$ and $\Phi (B,T,\mu)$ are obtained by
solving the system of two coupled equations that minimize
$\Omega_{B,T,\mu}^{\rm MFA, reg}$, viz.
\begin{align}
\frac{\partial \Omega_{B,T,\mu}^{\rm MFA, reg}}{\partial \bar{\sigma}} = 0\ ,
 & \qquad    \frac{\partial \Omega_{B,T,\mu}^{\rm MFA, reg}}{\partial \Phi} = 0\ .
\label{gap}
\end{align}
In addition, from the expression for the
thermodynamic potential in Eq.~(\ref{granpotMFA}) one can easily derive all
other relevant quantities. An important magnitude to be considered is the
(regularized) quark-antiquark condensate for each flavor, which is given by
\begin{equation}
\langle \bar q_f q_f \rangle_{B,T,\mu}^{\rm reg}=\frac{\partial \Omega_{B,T,\mu}^{\rm MFA, reg}}{\partial {m_{c}}}\ .
\label{average}
\end{equation}
In order to compare our results with those obtained from LQCD calculations,
it is useful to define the normalized flavor average condensate
\begin{equation}
 \bar \Sigma_{B,T}  = \frac{1}{2}(\Sigma^u_{B,T} + \Sigma^d_{B,T})\ ,
\end{equation}
where
\begin{equation}
 \Sigma^f_{B,T} = - \frac{2 m_c}{S^4}  \left[\langle \bar q_f q_f \rangle_{B,T,\mu}^{\rm reg} - \langle \bar q_f q_f \rangle_{0,0,0}^{\rm reg}
 \right]\ ,
\end{equation}
$S$ being a phenomenological scale fixed to $S = (135 \times 86)^{1/2}$~MeV~\citep{Bali:2012zg}.

\section{numerical results}
To obtain definite numerical results for the physical quantities of
interest, we need to specify the shape of the nonlocal form factor $g(p^2)$
in Eq.~(\ref{funcg}) and the model parameters $m_c$ and $G$. For simplicity
we consider a Gaussian form factor
\begin{equation}
g(p^{2}) = \exp\left( -p^{2}/\Lambda^{2} \right)\ ;
\end{equation}
this allows us to carry out the integral in Eq.~(\ref{funcg}) analytically,
leading to a significant reduction of computation time in subsequent
numerical calculations. The effective constituent quark masses in
Eq.~(\ref{masa_const}) are then given by~\citep{dumm2017strong}
\begin{align}
M_{k,p_\parallel}^{\pm,f} = \ & (1-\delta_{k,-1})\, m_{c} \nonumber \\
&+
\bar{\sigma}\frac{\left( 1-|q_{f}B|/\Lambda^{2}\right)^{k_\pm}}{\left(1+|q_{f}B|/\Lambda^{2}\right)^{k_\pm+1}}
\,\exp(-p_\parallel^{2}/\Lambda^{2})\ .
\end{align}
In any case, on the basis of previous analysis within this type of model, we
do not expect that the results show significant qualitative changes with the
form factor shape~\citep{dumm2017strong}. On the other hand, notice that the form factor introduces
a new dimensionful parameter $\Lambda$, which can be understood as an
effective soft momentum cutoff scale.

%In the present work, we use the set of parameters $m_{c}=6.5$ MeV, $\Lambda=678$ MeV and $G\Lambda^{2}=23.66$ \citep{dumm2017strong}. They were obtained by fixing the empirical values $m_{\pi}=139$ MeV and $f_{\pi}=92.4$ MeV and (at $\mu=T=eB=0$) for pion mass and weak decay constant, respectively, and a phenomenological value of chiral quark condensate $(-\langle{\bar{q}q}\rangle) ^{1/3}=230$  MeV (we call it P230) that better fits LQCD results (see Fig. 1 of Ref. \citep{dumm2017strong}).

To fix the free parameters $m_c$, $G$ and $\Lambda$ we demand that the
model reproduce the empirical values of the pion mass and decay constant,
$m_{\pi}=139$~MeV and $f_{\pi}=92.4$~MeV for vanishing $T$, $\mu$ and $B$.
In addition, we fit the parameters to a phenomenologically acceptable value
of the chiral quark-antiquark condensate, namely $200$~MeV~$< -\langle{\bar{q}q}\rangle^{1/3} < 300$~MeV. Some representative
parameter sets are quoted in Table~\ref{tab:sets}.
%Particularly for $(-\langle{\bar{q}q}\rangle) ^{1/3}=230$~MeV (P230), we obtain $m_{c}=6.5$ MeV, $\Lambda=678$ MeV and $G\Lambda^{2}=23.66$ \citep{dumm2017strong}.

\begin{table}[h]
\centering
\begin{tabular}{|c c c c c|}
\hline
\rule[-1ex]{0pt}{2.5ex} & \ \ \ $-\langle{\bar{q}q} \rangle^{1/3}$ \ \ \ & $m_c$ & $\Lambda$ & $G \Lambda^2$  \\
\hline
\hline
\rule[-1ex]{0pt}{2.5ex} P200 & 200 & 9.78 & 460 & 71.1 \\
\rule[-1ex]{0pt}{2.5ex} P230 & 230 & 6.49 & 678 & 23.7 \\
\rule[-1ex]{0pt}{2.5ex} P260 & 260 & 4.57 & 903 & 17.53 \\
\rule[-1ex]{0pt}{2.5ex} P300 & 300 & 3.03 & 1220 & 15.14 \\
\hline
\end{tabular}
 \caption{Some representative model parameter sets. Values of
$-\langle{\bar{q}q}\rangle^{1/3}$, $m_c$ and $\Lambda$ are given in MeV.}
\label{tab:sets}
\end{table}

\begin{figure}[htb]
%\centering
\includegraphics[width=0.4\textwidth]{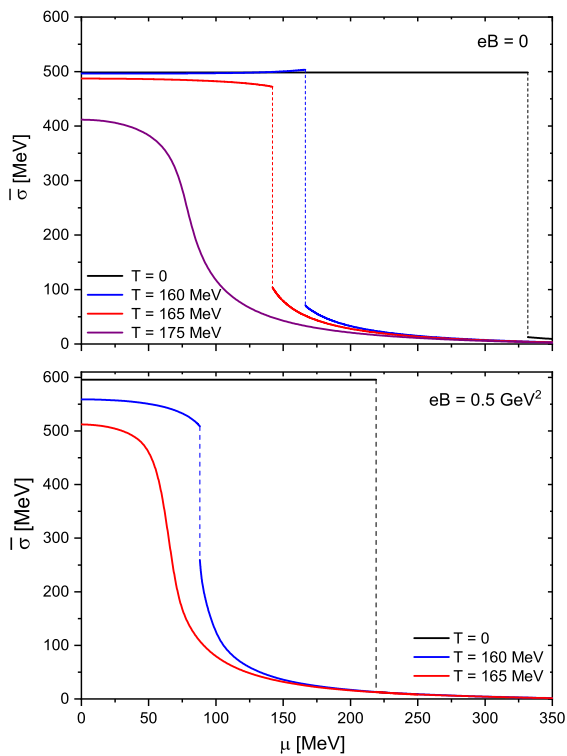}
\caption{Behavior of $\bar{\sigma}$ as a function of $\mu$ for different
values of the temperature and the magnetic field.}
\label{fig1}
\end{figure}

Once the parameter set has been chosen, we can solve numerically the gap
equations, Eq.~(\ref{gap}), for definite values of $T$, $\mu$ and $B$. As
expected, there are regions in which, given the magnetic field, there is
more than one solution for each value of $T$ and $\mu$. As usual we consider
that the stable solution is the one corresponding to the absolute minimum of
the potential. In Fig.~\ref{fig1} we show the behavior of $\bar{\sigma}$ as
a function of the quark chemical potential $\mu$ for $eB=0$ (upper panel)
and $eB=0.5$~GeV$^{2}$ (lower panel), taking several values of the
temperature. It is seen that for low values of $eB$ the chiral restoration
proceeds as a first order phase transition for temperatures up to about
$170$~MeV, while beyond this ``endpoint'' temperature the transition occurs
through a smooth crossover. If the magnetic field is increased, the first
order transition region gets reduced: the critical chemical potential for
the $T=0$ first order transition is found to be significantly lowered, and
the endpoint moves to lower values of $T$.

\begin{figure}[ht!]
\includegraphics[width=0.4\textwidth]{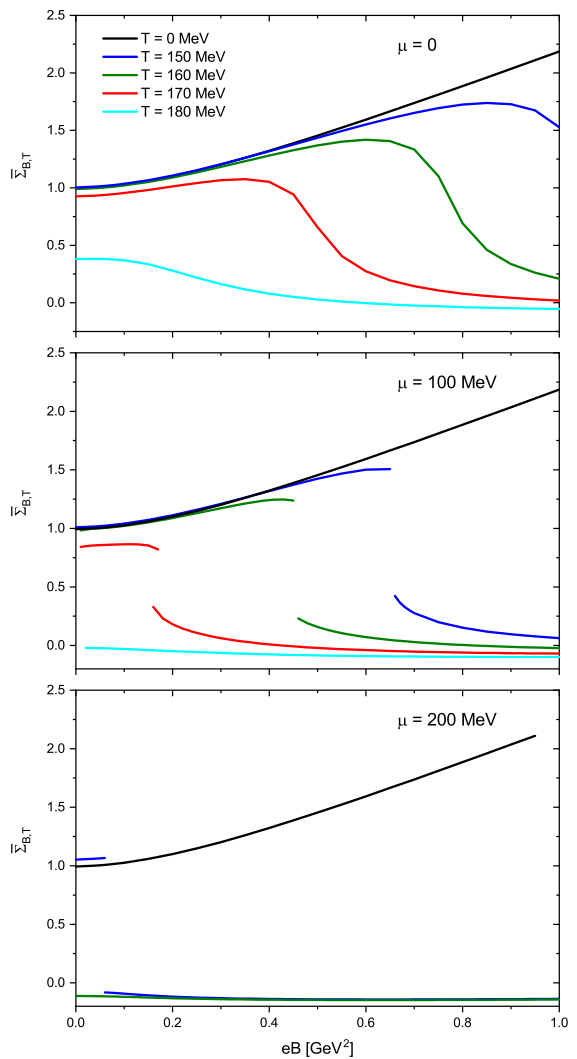}
\caption{Flavor-averaged chiral condensate as a function of $eB$ for
different representative values of the temperature and the quark chemical
potential.}
\label{fig2}
\end{figure}

As is well known, the chiral quark-antiquark condensates are
appropriate order parameters for the chiral symmetry restoration
transition. In Fig.~\ref{fig2} we show the behavior of the averaged chiral
condensate as a function of the external magnetic field, for some
representative values of both $T$ and $\mu$. The numerical
results correspond to parametrization P230. In all three panels it is seen
that for $T = 0$ the condensates show a monotonic increase with $B$, i.e.\
one has magnetic catalysis. On the contrary, when the temperature is
increased, for $\mu = 0$ the curves become non-monotonic, showing inverse
magnetic catalysis. This is also reflected in the fact that the chiral
restoration critical temperature gets reduced when the magnetic field is
increased~\cite{dumm2017strong}. Now, for values of the quark chemical
potential beyond $\sim 100$~MeV (see central and lower panels of
Fig.~\ref{fig2}) the system enters the first order transition region, and
the curves show discontinuities at some critical values of the magnetic
field.

To specify the critical temperatures for the crossover transitions we take
into account the maxima of the chiral susceptibility, defined as $\chi_{\rm
ch}=-\partial[(\langle{\bar{u}u}\rangle_{B,T,\mu}^{\rm reg} +
\langle{\bar{d}d}\rangle_{B,T,\mu}^{\rm reg})/2]/\partial T$. On the other
hand, as stated, for the deconfinement we take the Polyakov loop as the
relevant order parameter; therefore, to characterize the crossover-like
transitions we consider the PL susceptibility, which is defined as
$\chi_{\Phi} = \partial \Phi / \partial T$. It is seen that in the region
where both chiral restoration and deconfinement transitions proceed as
smooth crossovers, the peaks of the corresponding susceptibilities occur at
approximately the same temperatures and chemical potentials (i.e., both
transitions take place simultaneously).
\begin{figure}[ht]
\includegraphics[width=0.40\textwidth]{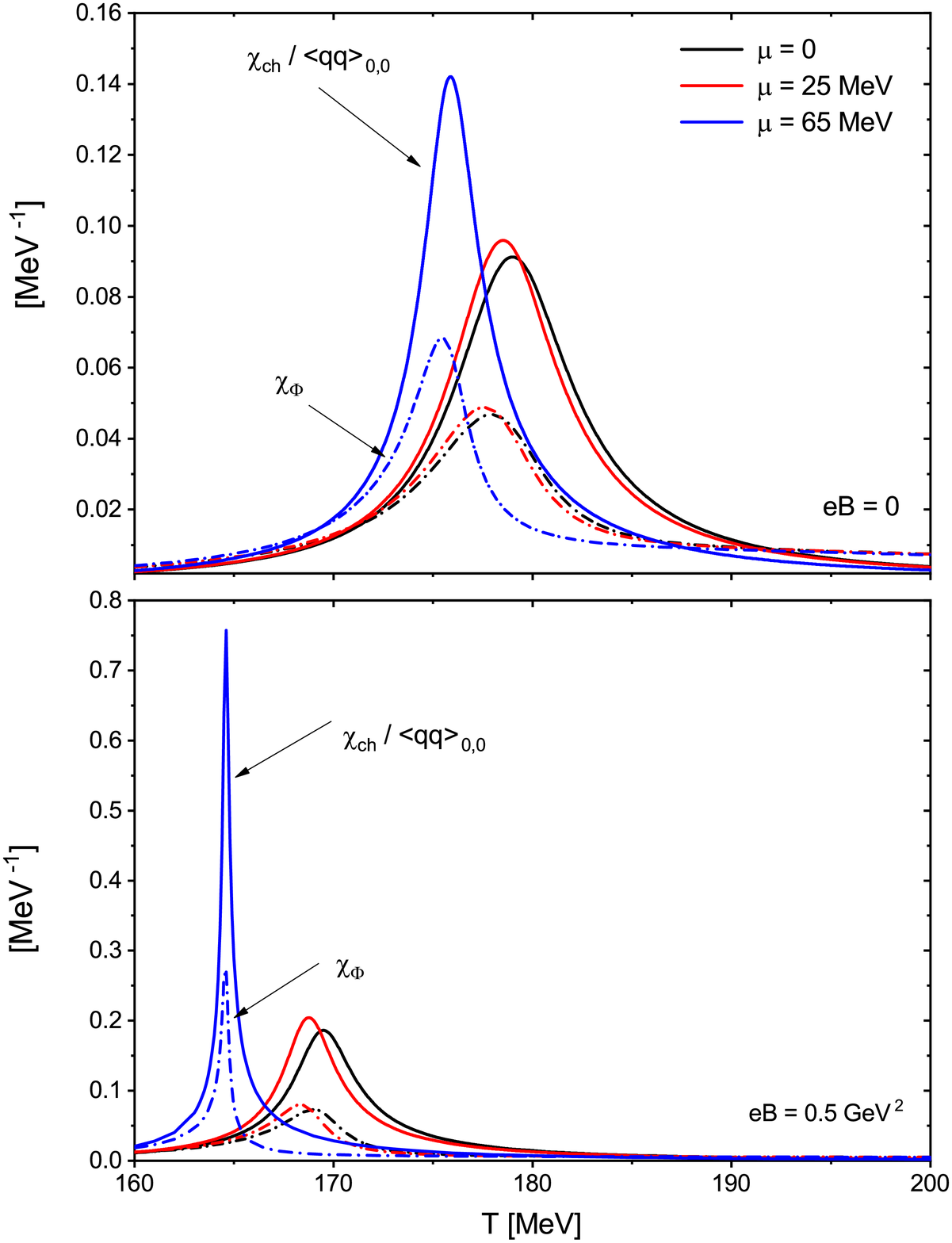}
\caption{Behavior of the chiral and PL susceptibilities as a function of
temperature for $eB=0$ and $eB=0.5$ GeV$^{2}$.  }
\label{fig4}
\end{figure}
This is shown in Fig.~\ref{fig4}, where we show the behavior of $\chi_{\rm
ch}$ and $\chi_\Phi$ for two different magnetic field strengths and some
representative values of the chemical potential corresponding to the
crossover region. As already mentioned, we see that the peaks in $\chi_{\rm
ch}$ and $\chi_{\Phi}$ are practically overlapped in the whole crossover
region,  i.e.\ the both crossover transitions are found to occur
simultaneously.

\begin{figure}[ht]
{\includegraphics[width=0.4\textwidth]{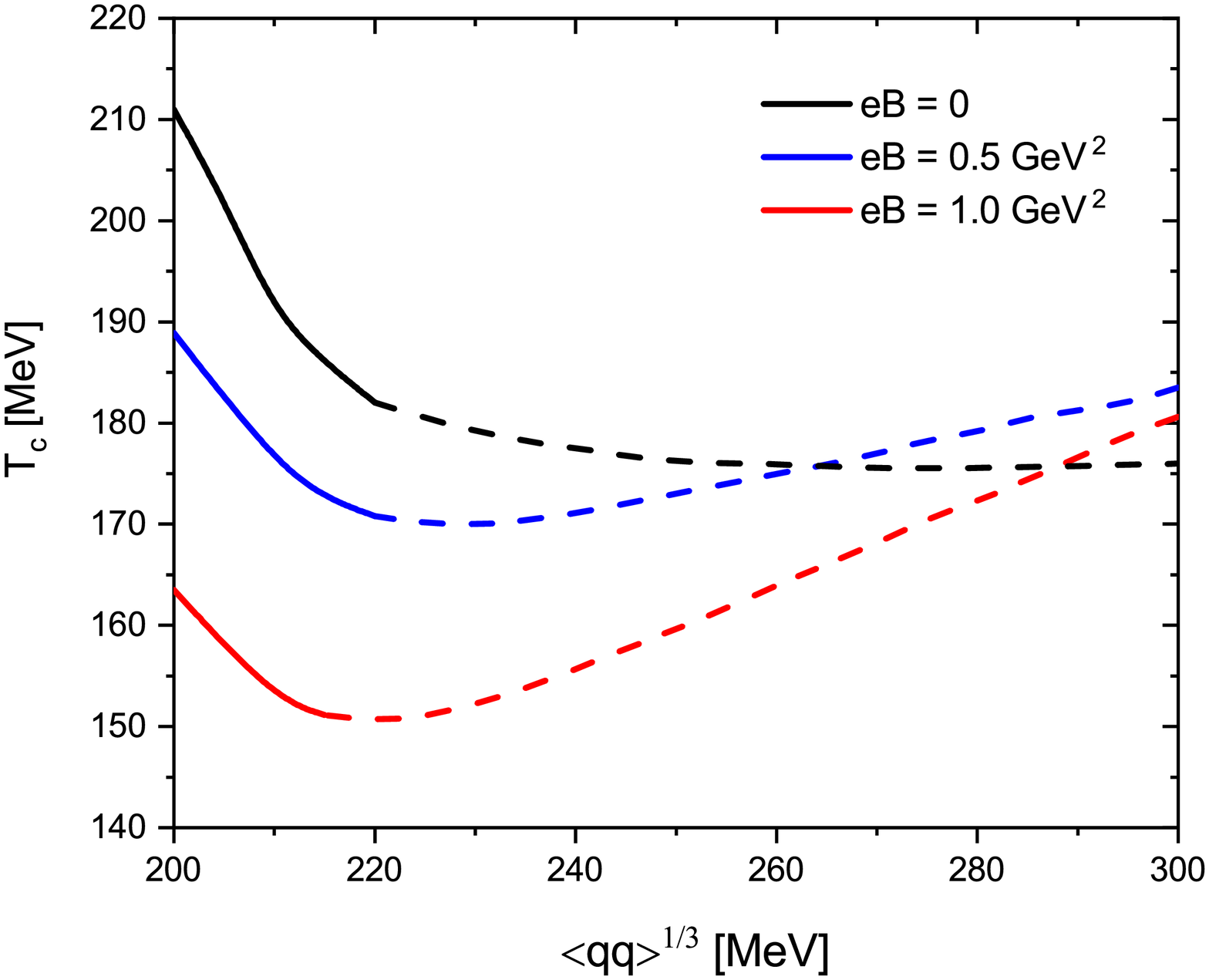}}\\
\vspace{0.25cm} {\includegraphics[width=0.4\textwidth]{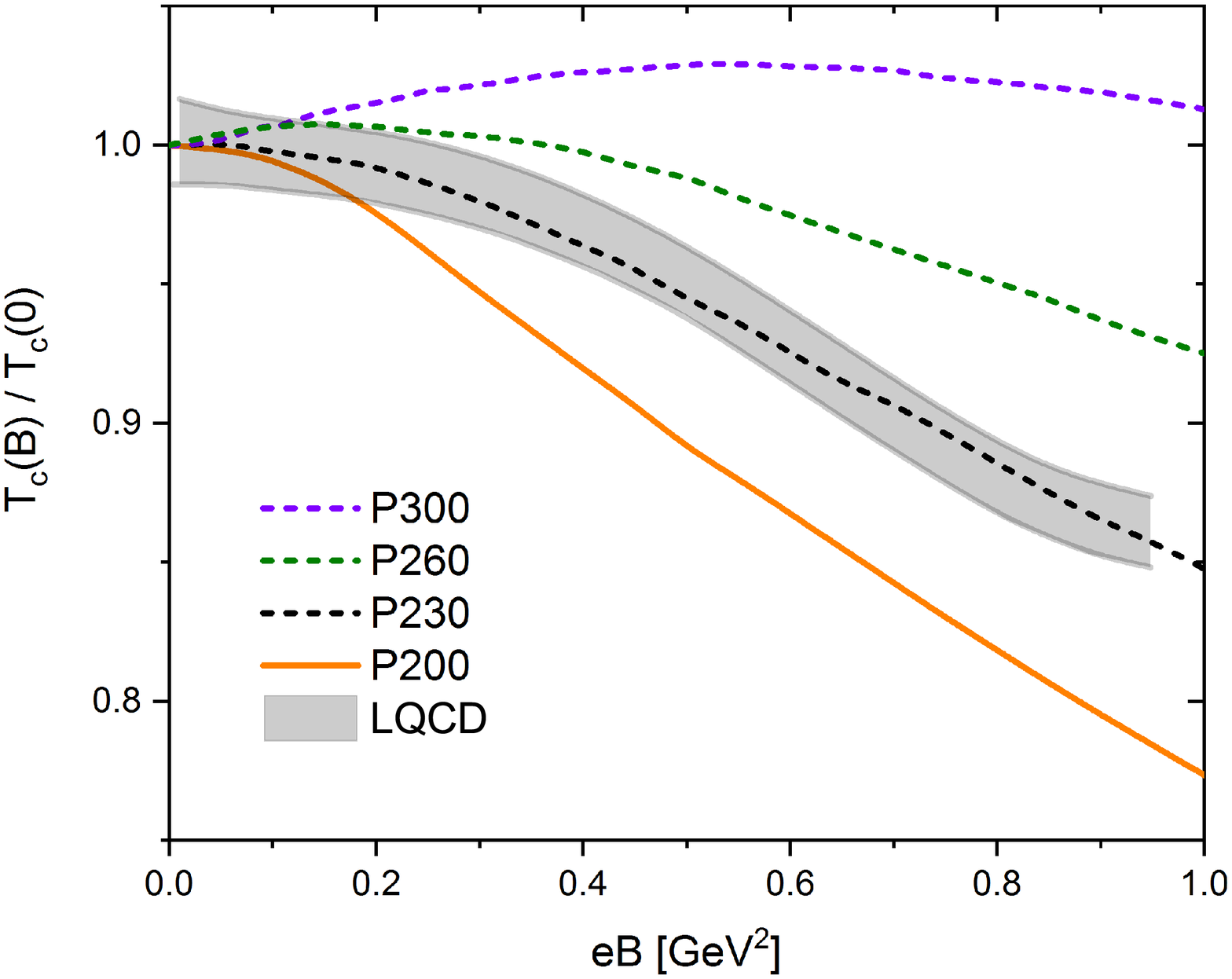}}
\caption{Upper panel: chiral critical temperatures $T_c$ at $\mu=0$ as functions of
the predicted value for the chiral quark condensate $-\langle
qq\rangle^{1/3}$, for different values of $eB$. Lower panel: Normalized critical
temperatures as functions of $eB$ for various model parametrizations. Solid
(dashed) lines correspond to first order (crossover) transitions. For
comparison, three-flavor LQCD results from Ref.~\cite{Bali:2012zg} are indicated by the
gray band.}
\label{fig3}
\end{figure}

It is also interesting to analyze the behavior of the critical chiral
restoration temperature for different sets of model parameters. In the upper
panel of Fig.~\ref{fig3} we show the critical temperature at $\mu=0$ as a
function of the parameter set, which we characterize by the value of the
$T=0$ chiral quark-antiquark condensate. We illustrate the situation by
considering three representative values of the external magnetic field. In
the figure, solid and dashed lines correspond to first-order and
crossover-like transition regions, respectively. Let us consider
firstly the $B=0$ case. It is seen that for parameter sets leading to
condensate values $-\langle \bar qq \rangle^{1/3} \alt 220$~MeV the
transition is predicted to be of first order, in contradiction with LQCD
calculations; on the other hand, for larger condensates one finds
crossover-like transitions at a stable critical temperature of about 175 to
180~MeV, in reasonable agreement with LQCD results. Going to the case of nonzero external magnetic
field, the blue and red lines correspond to $eB=0.5$ and $eB=1$~GeV$^2$,
respectively. It can be seen that for parametrizations leading to condensate
values up to 250~MeV the critical temperatures get reduced when the magnetic
field is increased, i.e.\ one observes the IMC effect; for larger
condensates, instead, the critical temperature reaches a maximum at some
value of $eB$. These behaviors are shown in the lower panel of
Fig.~\ref{fig3}, where we display the critical temperatures ---normalized to
the corresponding values at vanishing external magnetic field--- as
functions of $eB$, for different parametrizations. For P200 (solid line),
one has a first order transition for all values of the magnetic field. For
P230 the transition is crossover-like, and one has IMC; moreover, in this
case the critical temperatures are found to be compatible with the results
from LQCD quoted in Ref.~\cite{Bali:2012zg}, indicated by the gray band. Notice that LQCD calculations correspond to a three-flavor model; however, the normalized values of the critical temperatures are not expected to be significantly affected by the inclusion of strangeness. In
the case of P260 the IMC effect is reduced (in fact, the critical
temperature reaches a maximum for $eB\simeq 0.1$~GeV$^2$) and
gets progressively lost for parametrizations leading to larger values of the
chiral condensate. In view of these results we take P230 as our preferred
parametrization throughout this work.

\begin{figure}[ht]
{\includegraphics[width=0.4\textwidth]{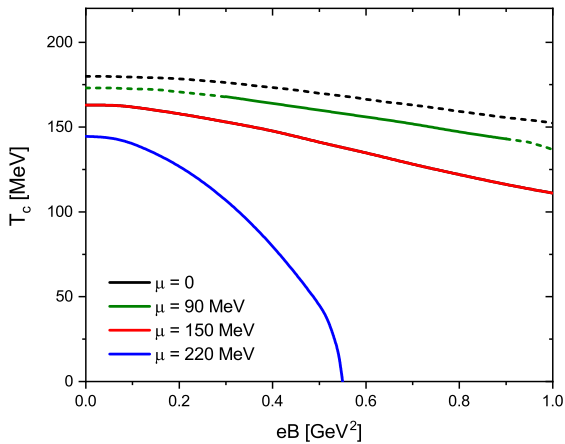}}
\caption{Chiral restoration critical temperature as a function of $eB$ for
some representative values of the quark chemical potential.}
\label{fig5}
\end{figure}

\begin{figure}[htb]
%\centering
\includegraphics[width=0.4\textwidth]{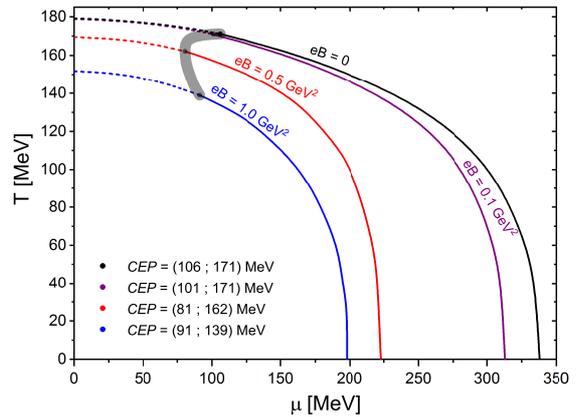}
\caption{Phase diagram in $\mu-T$ plane for different values of $eB$. Solid
and dashed lines correspond to first order and crossover transitions,
respectively. The gray band indicates the locations of the critical end
points.} \label{fig7}
\end{figure}

Next, in Fig.~\ref{fig5} we plot the critical temperature as a function of
the magnetic field, for some representative values of the quark chemical
potential. Starting from the curve for $\mu=0$ (black dashed line,
corresponding to the black dashed line in the lower panel of
Fig.~\ref{fig3}), it is seen that the IMC effect is preserved for nonzero
values of $\mu$. As expected, for relatively large values of $\mu$ the
curves reach the $T=0$ axis, since chiral symmetry is approximately restored
even at vanishing temperature (in absence of the external magnetic field
one has $\mu_c(T=0)\simeq 340$~MeV; no chiral symmetry broken
phase exists for values of $\mu$ beyond this limit~\citep{Ferraris_2021}).
Once again, solid (dashed) lines correspond to first order (crossover-like)
transitions. While for $\mu=0$ it is seen that the transition is smooth for
all values of $eB$, it becomes of first order for large values of $\mu$. In
the case of the curve corresponding to $\mu=90$~MeV (green line), we find a
first order piece for intermediate magnetic fields, viz.\ $0.3\alt eB\alt
0.9$. This is a manifestation of the particular behavior of the position of
CEP, and can be clearly seen from the $\mu-T$ phase diagram (see
Fig.~\ref{fig7}).

In Fig.~\ref{fig7} we show the phase diagram in the $\mu-T$ plane, taking
into account some values of the magnetic field to cover the range $eB=0$ to
1~GeV$^{2}$. As a general feature, it is seen that the pattern of a first
order transition line (solid) and a crossover transition line (dashed) that
meet at a critical end point is maintained for the considered range of
values of the magnetic field. As stated, the dashed lines correspond both to
the chiral restoration and the deconfinement transitions, which are found to
be strongly correlated. In addition, it is seen that crossover transition
lines get approximately overlapped for low magnetic fields ($0\leq eB \alt
0.1$~GeV$^{2}$). It is worth looking at the location $(\mu_{\rm CEP},T_{\rm
CEP})$ of the critical end point for the studied magnetic field range; the
corresponding values are indicated in the figure by the wide gray line. It
is found that the temperature $T_{\rm CEP}$ decreases steadily with the
magnetic field, while the chemical potential $\mu_{\rm CEP}$ lies between
about 80 to 105 MeV, reaching its minimum for intermediate values of the
magnetic field. This behavior is remarkably different from the one obtained
in Refs.~\citep{Avancini:2012ee,Ferrari:2012yw} in the framework of 2- and
3-flavor local NJL-like models, where the CEP temperature is found to grow
when the external magnetic field get increased. Our results also differ
qualitatively from those obtained in Ref.~\citep{Costa:2015bza}, where the
authors consider a PNJL model in which the IMC effect is reproduced by
including a $B$-dependent four-quark coupling.

\begin{figure}[htb]
\centering
\includegraphics[width=0.4\textwidth]{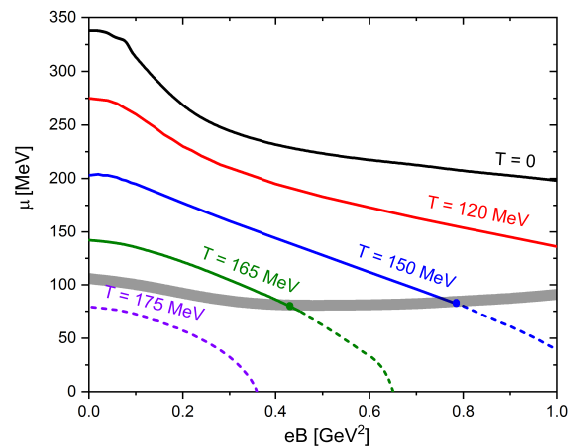}
\caption{Phase diagram in $\mu-eB$ plane for different temperatures. Solid
and dashed lines represent first-order and crossover transitions,
respectively. The gray band indicates the locations of the critical end
points.} \label{fig6}
\end{figure}

Finally, for completeness we show in Fig.~\ref{fig6} the phase
diagram in the $\mu-eB$ plane, considering several representative values of
the temperature. The curve for $T=0$, previously obtained in
Ref.~\cite{Ferraris_2021}, shows that the critical chemical potential is a
decreasing function of the magnetic field, and the transition is of first
order for the considered range of values of $B$. We see from the figure that
this decreasing behavior is maintained for larger values of $T$, while the
curves reach the CEPs (indicated by the wide gray line) at $B<1$~GeV$^2$ for
temperatures below $\simeq 140$~MeV. For values of $T$ between $\simeq 170$
and 180~MeV only crossover transitions can occur, whereas (as one can see
from Fig.~\ref{fig7}) no chiral symmetry broken phase exists for
temperatures above $T_c(B=0)=180$~MeV.

\section{Summary and conclusions}

We have studied the QCD phase diagram, for both nonzero temperature and
chemical potential, in the presence of an external static and uniform magnetic field.
Our analysis has been done in the framework of a two-flavor nonlocal version
of the NJL model, including the couplings of fermions to the Polyakov loop.
This model has the feature of offering a natural mechanism for the
description of the IMC effect observed through LQCD calculations.

Our numerical analysis shows that the IMC, reflected by a decreasing
behavior of the critical chiral transition temperature as a function of the
magnetic field, is preserved for nonzero values of the chemical potential.
In addition, the qualitative features of the transition curves known for
$B=0$ ---consisting in a crossover region and a first order transition line,
separated by a critical end point--- is maintained for values of $eB$ ranging from zero to 1~GeV$^2$. From the analysis of the behavior of the traced Polyakov loop, it is seen that in the crossover region chiral restoration and deconfinement transitions occur simultaneously, in agreement with LQCD predictions. Moreover, it is found that the CEP exists for the whole considered range of the magnetic field, in a region that could be accessible for relativistic heavy ion collision experiments. 

Our results for the location $(\mu_{\rm CEP},T_{\rm CEP})$ of the critical end point show that the chemical potential $\mu_{\rm CEP}$ lies in a relatively narrow region, between say 80 and 105~MeV, reaching the minimum for an intermediate value of the magnetic field, $eB\simeq 0.5$~GeV$^2$. On the other hand, the temperature $T_{\rm CEP}$ is found to decrease monotonously with the magnetic field, from $T_{\rm CEP}\simeq 170$~MeV for $eB=0$ to $T_{\rm CEP}\simeq 140$~MeV for $eB\simeq 1$~GeV$^2$. This behavior is opposite to the one observed in most studies of local versions of the NJL/PNJL models, in which $T_{\rm CEP}$ gets enhanced for increasing values of the magnetic field~\citep{Avancini:2012ee,Ferrari:2012yw,Costa:2013zca,Costa:2015bza,Ferreira:2017wtx}. 
We find it natural to attribute this qualitative difference to the fact that the nonlocal PNJL has the feature of showing IMC at vanishing chemical potential; as stated, the decrease of the critical temperature with the magnetic field also occurs at finite $\mu$, leading to the descent of the temperature of the CEP. In fact, in the local NJL model it is also found that the behavior of $T_{\rm CEP}$ with the magnetic field is significantly modified if one considers a $B$-dependent effective quark coupling such that the model could show IMC~\citep{Costa:2015bza}. It is also worth mentioning that although our results have been obtained for a two-flavor model, the values of $\mu_{\rm CEP}$ remain well below the $s$ quark threshold, and we have verified that the IMC observed at finite $\mu$ is robust under moderate changes in the model parameters; hence, the behavior of the CEP we have found in the low density region should not be qualitatively modified by the inclusion of strangeness degrees of freedom  (notice that a more complex structure, including other CEPs, could be found at higher densities~\cite{Ferreira:2017wtx}). In this way, although it is still difficult to get definite predictions about the CEP location, it is seen that its behavior as a function of the magnetic field can be taken as an important clue to get insight on the character of effective four-point quark interactions and the effect of inverse magnetic catalysis. 

\begin{acknowledgments}
We are grateful to N.N. Scoccola for useful discussions and a critical reading of the manuscript. J.P.C, D.G.D and A.G.G. acknowledge financial support from CONICET under
Grant No.\ PIP 22-24 11220210100150CO, ANPCyT (Argentina) under Grant
PICT20-01847 and PICT19-00792, and the National University of La Plata
(Argentina), Project No.\ X824.
\end{acknowledgments}

\bibliographystyle{apsrev4-1}
\bibliography{TMUB}

\end{document}